# MRI quantification of liver fibrosis using diamagnetic susceptibility: An ex-vivo feasibility study.

Chao Li[1,2], Jinwei Zhang[1,3], Alexey V. Dimov[1], Anne K. Koehne de González[4], Martin R. Prince[1], Jiahao Li[1,3], Dominick Romano[1,3], Pascal Spincemaille[1], Thanh D. Nguyen[1], Gary M. Brittenham[5], and Yi Wang[1]

1. Radiology, Weill Cornell Medicine, New York, NY, United States
2. Applied and Engineering Physics, Cornell University, Ithaca, NY, United States
3. Meinig School of Biomedical Engineering, Cornell University, Ithaca, NY, United States
4. Department of Pathology, Columbia University, New York, NY, United States
5. Department of Pediatrics, Columbia University, New York, NY, United States

Abstract (298/300 words)

In chronic liver disease, liver fibrosis develops as excessive deposition of extracellular matrix macromolecules, predominantly collagens, progressively form fibrous scars that disrupt the hepatic architecture, and fibrosis, iron, and fat are interrelated. Fibrosis is the best predictor of morbidity and mortality in chronic liver disease but liver biopsy, the reference method for diagnosis and staging, is invasive and limited by sampling and interobserver variability and risks of complications. The overall objective of this study was to develop a new non-invasive method to quantify fibrosis using diamagnetic susceptibility sources with histology validation in ex vivo liver explants. In 20 formalin-fixed liver explant sections, histological examination provided semiquantitative staging of fibrosis, iron, and steatosis. Fibrous extracellular matrix components, primarily fibrillar collagens, are diamagnetic while iron is paramagnetic. From multi-echo gradient echo (mGRE) data, proton density fat fraction (PDFF), QSM, $R2^*$, and $R2^*$ based negative susceptibility maps were generated. $R2^*$ based susceptibility distinguished samples with no or mild fibrosis (stages F0 to F1) from moderate to advanced fibrosis (stages F2 to F3; p=0.0025), and stages F2 to F3 from cirrhosis (stage F4; p=0.021), and provided an 0.987 (p=0.0017) area under the curve (AUC) of receiver operating characteristic to differentiate between samples with no, mild or moderate fibrosis (stages F0 to F1) and those with significant fibrosis or cirrhosis (stages F2 to F4) with a sensitivity of 93%, a specificity of 100%. This diagnostic utility of $R2^*$ based negative susceptibility was superior to that of R2 derived from multi-echo spin echo data, $R2^*$, $R2'=R2^*-R2$, $R2'$ based negative susceptibility, or QSM. These results provide evidence that negative susceptibility modeled from QSM and $R2^*$ could improve the MRI staging of fibrosis in chronic liver disease.

# 1 | INTRODUCTION

Worldwide, chronic liver disease affects 1.5 billion persons, causing two million deaths each year.[1,2] The most common etiologies are i) nonalcoholic fatty liver disease (NAFLD) (or, as has been proposed,[3] metabolic-dysfunction associated steatotic liver disease, MASLD, including nonalcoholic steatohepatitis, NASH, or metabolic-dysfunctio associated steatohepatitis, MASH), ii) viral infections (hepatitis B virus, HBV, and hepatitis C virus, HCV), and iii) alcohol-related liver disease (ALD); other etiologies include less common genetic, autoimmune, inflammatory, metabolic, infectious, and toxic disorders. In chronic liver disease, fibrosis, iron, and fat are interrelated and iron may be both a cause[4] and consequence of liver disease.[5-7] Regardless of the etiology of chronic liver disease, the molecular makeup of the fibrous scar tissue is similar and consists of collagen types I and III, sulfated proteoglycans, and glycoproteins.[8] In chronic liver disease, fibrosis develops with excessive deposition of extracellular matrix macromolecules, predominantly collagens. This progressively forms fibrous scars that replace normal tissue and distort hepatic architecture. Cirrhosis develops following formation of nodules of regenerating hepatocytes, potentially leading to portal hypertension, liver failure and hepatocellular carcinoma.[9-11]

Fibrosis is the best predictor of morbidity and mortality in chronic liver disease.[12,13] Liver biopsy, the reference method for diagnosis and staging of hepatic fibrosis, is invasive, cannot be used for frequent, repeated monitoring, and is limited by sampling and interobserver variability, and by risks of complications. Consequently, alternative methods that avoid biopsy are needed for patient care and in clinical research.[14] The recent FDA approval of the selective thyroid hormone receptor β agonist resmetirom as the first drug for the treatment of adults with noncirrhotic non-alcoholic steatohepatitis (NASH or MASH), can be expected to increase the demand for non-invasive means to evaluate hepatic fibrosis.[15,16]

While a variety of approaches are being explored, quantitative imaging methods using MRI for diagnosis and staging offer unique advantages because of their ability to generate voxel-wise data that characterize intrinsic liver tissue properties.[14] Previous studies have examined multiparametric approaches using liver stiffness for fibrosis, T2/ R2, and T2*/R2* for iron, and proton density fat fraction (PDFF) for fat.[17-21] Quantitative susceptibility mapping (QSM) derives the magnetic susceptibility of tissues from the tissue field captured through complex multi-echo gradient echo (mGRE) data[22,23]. The extracellular matrix proteins and iron in a fibrotic liver are major susceptibility sources affecting mGRE signal magnitude and phase. The combined R2* and QSM can distinguish between different stages of fibrosis both in ex vivo liver explants[24] and in vivo.[25] More recently, methods for susceptibility source separation in the brain have been developed using R2'=R2*-R2 and QSM.[26] Here, we describe a new non-invasive method to quantify fibrosis in the liver using negative susceptibility determined by QSM from mGRE phase and R2* from mGRE magnitude. Referencing semiquantitative histological staging of fibrosis, we investigate the ability of R2* based susceptibility, R2'-based negative susceptibility, and the individual parameters R2, R2*, R2' and QSM to differentiate between samples with no, mild or moderate fibrosis and those with advanced fibrosis or cirrhosis.

# 2 | METHODS

## 2.1 | Liver samples

This study was approved by the Institutional Review Board. Between January 2021 and November 2023, liver explant samples were collected by a liver pathologist. Sections of the livers of ~7 x 5 x 1.5 cm³ size and ~40 grams weight were selected, preserved in formalin, and subsequently imaged with MRI.

## 2.2 | MRI acquisition

Liver samples preserved in formalin were placed in a cylindrical agarose mold covered with a water balloon. MRI was performed on a 3T scanner (GE Healthcare, Waukesha, WI, USA). Imaging sequences included 1) 3D mGRE with 8 echoes, flip angle = 15, TE1 = 2.6 ms, ΔTE = 2.7 ms, TR = 24.43 ms, reconstructed voxel size = 0.88 × 0.88 × 1 mm³, bandwidth = 390 Hz/pixel, reconstructed matrix = 256 × 256 x 74-128; 2) 2D multi-echo spin-echo (mSE) with 8 echoes, TE = 6.6, 13.2, 19.8 26.4, 33.1, 39.7, 46.3, and 52.9 ms, band-width = 244 Hz/pixel, FOV = 240 ×240 mm², voxel size = 0.9375 ×0.9375 mm² , slice thickness = 1 mm, number of slices = 18-32, TR = 1500 ms,.

## 2.3 | MRI analysis – R2, R2*, and fat water separation

MRI analysis was conducted using MATLAB (version R2023b; MathWorks, Natick, MA, USA). R2* and R2 values were derived from the magnitudes of the mGRE signals and mSE signals, respectively, using the ARLO method [27]. An initial estimation of the field f was produced from the complex mGRE signals S(t) with N echoes, achieved through concurrent phase unwrapping and chemical shift elimination (SPURS) [28]. The IDEAL algorithm [29] was applied to create maps of water ($W$), fat ($F$), and field ($f$) mGRE data from based on a single-peak fat model:

$$E(W,F,f) = \mathrm{argmin}_{W,F,f} \sum_{j=1}^{N} \left\| S(t_j) - e^{-R_2^* t_j} e^{-i2\pi f t_j}(W + F e^{-i2\pi v \cdot t_j}) \right\|_2^2 \quad (1)$$

Background field was removed using the Projection onto Dipole Fields (PDF) technique [30]. The local field map was then used to produce a susceptibility map (χ) through the Morphology Enabled Dipole Inversion (MEDI) algorithm [31-33]. The susceptibility measurements were referenced to the mean susceptibility value of the water balloon [32]. Images representing the Proton Density Fat Fraction (PDFF) were generated from the computed fat and water images as $F/(W + F)$ [34]. Next, the susceptibility maps were adjusted for fat content using the PDFF maps, based on the assumption that the susceptibility of pure fat is 0.65 ppm (18), as described in (19).

To exclude the background agarose gel and air, liver specimens volumes were automatically created by combining threshold-based segmentation with manual segmentation. This involved generating a liver binary mask $M_L$ using an R2*≥15 s⁻¹ threshold[24]. The mask was then manually eroded around large blood vessels, and any air bubbles were also excluded to prevent inaccuracies in fibrosis detection. Finally, the average values of all MRI parameters were calculated within the mask.

## 2.4 | MRI analysis – QSM source separation

QSM $\chi$ susceptibility at a voxel was regarded to comprise positive source $\chi^+$ and negative source $\chi^-$:

$$\chi = |\chi^+| - |\chi^-| \quad (2)$$

These positive and negative sources were assumed to contribute additively to the static dephasing rate $R_2'$ [35], which was the difference between $R_2^*$ determined from mGRE and $R_2$ from mSE [36]:

$$R_2' = R_2^* - R_2 = D_{\phantom{x}}^+|\chi^+| + D_{\phantom{x}}^-|\chi^-| \quad (3)$$

The dephasing constants $D_{\phantom{x}}^+ = D_{\phantom{x}}^- = D$ were assumed to be the same and may vary with tissues. Note that mSE sequence has different voxel size from mGRE, requires a long acquisition time, and is not commonly acquired in clinical MRI. R2' can be assumed to be proportional $\alpha$ to R2* from only mGRE data[37]:

$$R_2'(r) \approx \alpha R_2^*(r) \quad (4)$$

The positive and negative susceptibility maps were calculated by solving the optimization problem with regularization[36-38]

$$\chi^{+*}, \chi^{-*} = \arg \min_{\chi^+, \chi^-} \|w_1(R_2' - D(|\chi^+| + |\chi^-|))\|_2^2 + \|w_2(f - d * (\chi^+ + \chi^-))\|_2^2$$
$$+ 2\lambda_1|M_E \nabla(\chi^+ + \chi^-)|_1 + \lambda_1|M_L \nabla \chi^+|_1 + \lambda_1|M_L \nabla \chi^-|_1 \quad (5)$$
$$+ \lambda_2\|M_b(\chi^+ - \overline{\chi_b^+})\|_2^2 + \lambda_2\|M_b(\chi^- - \overline{\chi_b^-})\|_2^2$$

where $f$ is the local field as in Eq.1, $\lambda_i$ are regularization parameters, $\nabla$ is a gradient operator, $M_E$ is a binary edge mask derived from the magnitude image, $M_L$ is a binary edge mask derived from $R_2^*$, $M_b$ is a binary mask of the water balloon, and $\overline{\chi_b^+}$ and $\overline{\chi_b^-}$ are $\chi^+$ and $\chi^-$ averaged over, respectively. $w_1$ and $w_2$ are data weighting terms. Susceptibility values violating $\chi^+ > 0$ and $\chi^- < 0$ were reset to zero.

The dephasing constant was calculated in the brain [36] based on the average $R_2'$ values from five deep gray matter regions that were assumed to have only positive susceptibility source. However, such a region of single susceptibility source does not exist in the liver tissue, which everywhere is a variable mixture of diamagnetic fibrosis and paramagnetic iron and fat. In this work, we estimated the dephasing constant $D$ such that the obtained negative susceptibility best differentiated various fibrosis groups. This was achieved by minimizing the sum of the losses for the logistic models for F0-1 vs F2-3, F2-3 vs F4 and F0-1 vs F4:

$$\log \text{Loss} = -\sum_{j=1}^{3}\sum_{i=1}^{N_j} [y_i^j \log(p_i^j) + (1-y_i^j)\log(1-p_i^j)] \quad (6)$$

Here $j = 1$ stands for the logistic model differentiating F0-1 vs F2-3, $j = 2$ stands for the model differentiating F2-3 vs F4, and j = 3 for F0-1 vs F4. When a relaxometry constant D is given, an average negative susceptibility can be solved rapidly using Eqs 2 and 3. This value for each liver sample was used as the input of the logistic models to predict the labels of the sample. $y_i^j$ is the label for sample $i$ in model $j$ (For example, if $j = 1$, the model only contained data with F0-1 and F2-3), and $N_j$ is the total number of samples for this model. $p_i^j$ is the prediction of sample by model $j$ given the average negative susceptibility for sample $i$. We examined the values for $D$ from 50Hz/ppm to 200Hz/ppm with a step of 1Hz/ppm. After obtaining the optimal value for $D$, we generated the final susceptibility map using Eq.5.

In this study, we performed both R2'-based and R2*-based source separation to compare their ability in differentiating the fibrosis stages. To ensure a fair comparison, we solved the optimization problem (6) using R2' and R2* separately to get two different sets of dephasing constants $D$ and used them in their corresponding source separation procedures.

## 2.5 | Histopathological analysis

A small section of each scanned liver sample was collected for histopathological analysis for each case. Hematoxylin and eosin (H&E), Masson's trichrome, and Prussian Blue stains were performed for histology, fibrosis, and iron evaluation, respectively. A liver pathologist evaluated the sections and assigned scores for fibrosis, iron, and fat using standard clinical scoring systems [39-41]. Fibrosis appeared blue fibers on liver tissue samples stained with Masson's trichrome examined under a microscope. Iron appeared as blue microscopic granules within cells on Prussian blue staining.

## 2.6 | Statistical analysis

Differences in average R2*, R2, R2', $\chi$ and $|\chi^-|$ between the subgroups F0-1 (non-fibrotic, mild fibrotic), F2-3 (medium stage fibrotic) and F4 (cirrhotic) and between the subgroups F0-2 (non-fibrotic, mild fibrotic and significant stage fibrotic) and F3-4 (advanced stage fibrotic) liver samples were evaluated using Mann–Whitney U tests.

Since NASH patients with ≥F2 are the target population for pharmacological treatment and are most likely to benefit from antifibrotic drugs [42], the diagnostic accuracy of individual parameters for fibrosis detection between subgroups of liver samples with stages F0-1 and F2-4 was evaluated through Receiver Operating Characteristic (ROC) curve analysis. The optimal diagnostic threshold was determined using the Youden index, from which sensitivity and specificity values were calculated at this threshold.

# 3 | RESULTS

## 3.1 | Demographics and histopathological characteristics

A total of 22 patient samples underwent scanning, with 20 of these being subjected to analysis. Two samples were excluded from the analysis due to the failure of water-fat separation and QSM reconstruction, attributed to extremely high R2* values caused by iron overload. These excluded samples only displayed a discernible liver signal in the initial echo of mGRE images.

Table 1 presents the demographics and histopathological characteristics across the entire cohort and the different fibrosis groups (F0-1 vs. F2-3 vs. F4). Among the 20 samples analyzed, 8 displayed liver cirrhosis (stage 4), 2 indicated stage 3 fibrosis, 5 were identified with stage 2 fibrosis, and 2 with stage 1 fibrosis. The other 3 samples did not present any fibrosis signs, leading to a classification of 7 samples as intermediate-stage fibrosis (F2-3, combining stages 2 and 3) and 5 as either non-significant fibrosis or non-fibrotic (F0-1, combining stages 0 and 1). Iron deposition, noted either in hepatocytes or Kupffer cells, was observed in 10 samples, with 7 belonging to the F4 group, 2 to the F2-3 group, and 1 to the F0-1 group, indicating a noticeable trend towards a higher prevalence of iron deposition among the cirrhotic samples. Steatosis exceeding 5% was observed in 6 samples, including 3 from the F4 group, 1 from the F2-3 group, and 2 from the F0-1 group. Notably, 4 samples within the cirrhotic group exhibited minimal steatosis (less than 5%).

**TABLE 1** Demographic and histopathologic characteristics of the liver samples.

| | All (n = 20) | F0-1 (n = 5) | F2-3 (n = 7) | F4 (n = 8) |
|---|---|---|---|---|
| Age (y) | 43 (0-80) | 13 (0-80) | 45 (24-61) | 57 (49-66) |
| Sex | | | | |
| -Male | 13 | 4 | 4 | 5 |
| -Female | 7 | 1 | 3 | 3 |
| Iron deposition | | | | |
| -None | 10 | 4 | 5 | 1 |
| -Grade 0 | 1 | 0 | 1* | 0 |
| -Grade 0-1 | 1 | 0 | 1 | 0 |
| -Grade 1 | 1 | 1 | 0 | 2 |
| -Grade 2 | 1 | 0 | 0 | 1 |
| -Grade 2-3 | 3 | 0 | 0 | 3 |
| -Grade 3 | 1 | 0 | 0 | 1 |
| Steatosis | | | | |
| -None | 9 | 2 | 6 | 1 |
| -Stage 0 (<5%) | 5 | 1 | 0 | 4 |
| -Stage 1 (5-33%) | 4 | 1 | 1 | 2 |
| -Stage 2 (34-66%) | 1 | 1 | 0 | 0 |
| -Stage 3 (>66%) | 1 | 0 | 0 | 1 |

*In one sample where grade-0 iron was found in the hepatocytes, iron was observed in the Kupffer cells.

### 3.2 | $\chi^-$, R2*, R2, R2' and QSM measurements

Figure 1 illustrates a linear relation between average R2' and R2* values across all the liver samples according to Eq.4: $\alpha = 0.76$ and $R^2 = 0.95$ indicating a good fit.

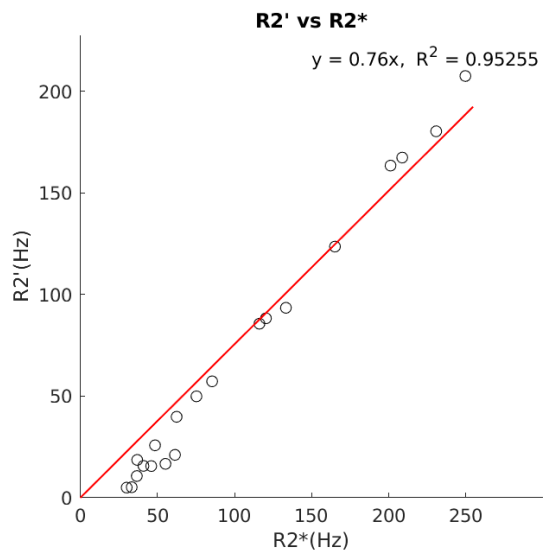

Figure 1: Estimation of proportionality constant $\alpha$ in Eq.4.

Figure 2 exemplifies QSM $\chi$ maps, R2* based $|\chi^+|$ and $|\chi^-|$ maps with $\alpha = 0.76$, and R2*, R2 and R2' maps. The optimal dephasing constant according to Eq.5 for distinguishing between fibrosis stages F0-1/F2-3 and F2-3/F4 was identified around 109 Hz/ppm for R2* and QSM data, and 125Hz/ppm for R2' and QSM data.

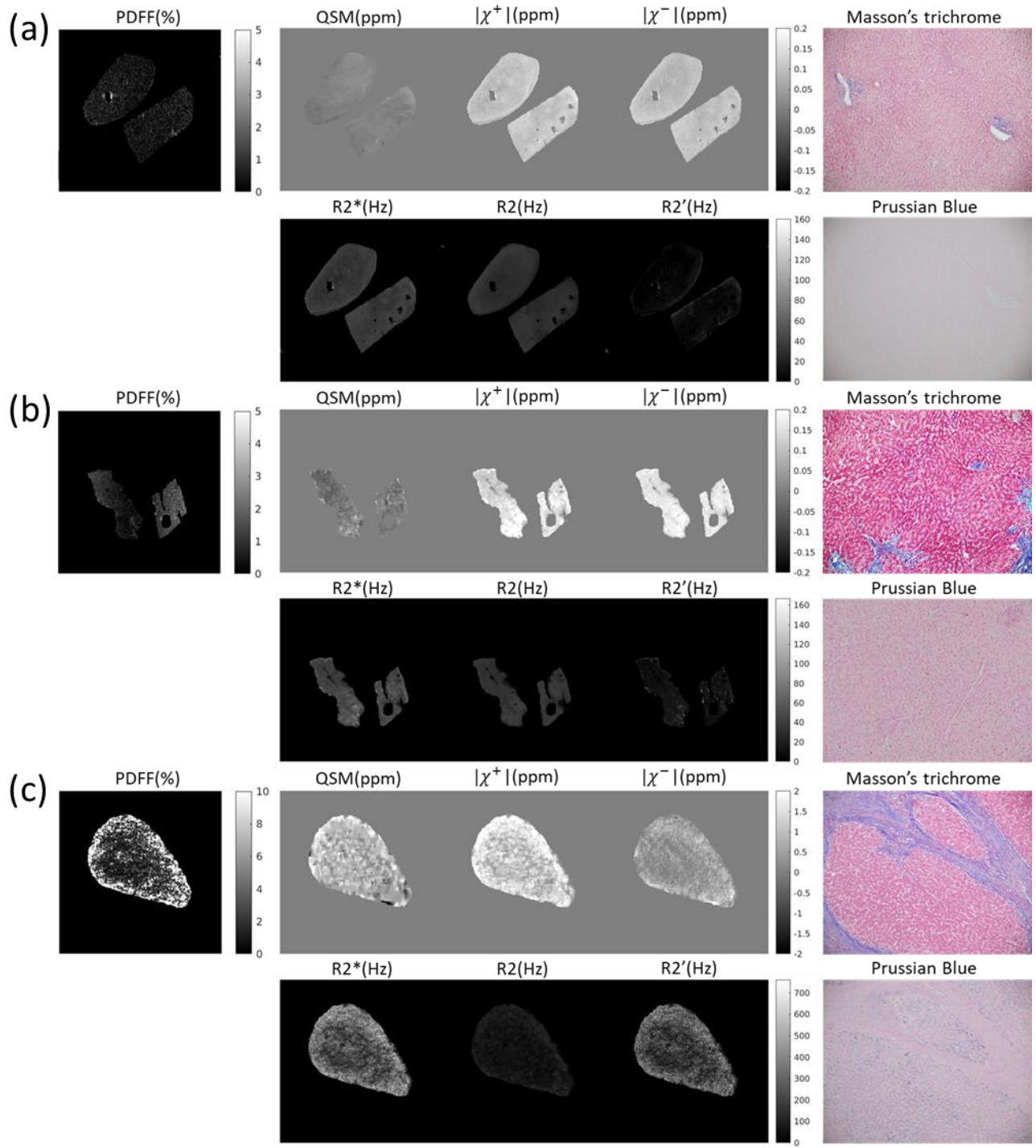

Figure 2: $\chi$, $|\chi^+|$ and $|\chi^-|$ maps (R2* base), R2*, R2 and R2' in a cross-sectional slice through a nonfibrotic liver sample (top), intermediate stage (F2) fibrosis case (middle) and a cirrhotic liver sample (bottom). Masson's trichrome and Prussian blue are shown for each example.

Figure 3 presents boxplots of R2* based absolute negative susceptibility $|\chi^-|$, R2' based $|\chi^-|$, R2*, R2, R2' and χ across the subgroups of F0-1, F2-3 and F4 of liver fibrosis. There was an increase in both the average R2'-based and R2* based $|\chi^-|$, when comparing non-fibrotic or mildly fibrotic samples

(F0-1, mean = 0.11 ± 0.015 for R2* based $|\chi^-|$ and mean = 0.033 ± 0.022 for R2' based $|\chi^-|$) to medium-stage fibrosis (F2-3, mean = 0.17 ± 0.037 for R2* based $|\chi^-|$ and mean = 0.075 ± 0.049 for R2' based $|\chi^-|$), and again from F2-3 to cirrhotic samples (F4, mean = 0.39 ± 0.23 for R2* based $|\chi^-|$ and mean = 0.34 ± 0.24 for R2' based $|\chi^-|$), suggesting that $\chi^-$ may correlate with the severity of fibrosis. The $|\chi^-|$ difference between F2-3 and F4 was significant for both R2* (p = 0.021) and R2' (p = 0.004) based susceptibility separation, but $|\chi^-|$ difference between F0-1 vs F2-3 was only for R2* (p = 0.0025) but not for R2' (p = 0.20) based susceptibility separation.

R2* values also showed a significant increase from F0-1 to F2-3 (p = 0.048) and from F2-3 to F4 (p = 0.021), implying its potential utility in distinguishing between fibrosis stages. R2' values was significantly different only between the F2-3 and F4 groups, and R2 value was only significantly different between F0-1 and F2-3. QSM was not statistically different across the groups, consistent with fibrosis and iron mixing across all stages.

Figure 4 shows boxplots of R2* based $|\chi^-|$, R2' based $|\chi^-|$, R2*, R2, R2', and χ across the subgroups of F0-2, F3-4 of liver fibrosis. R2'-based $|\chi^-|$, R2*-based $|\chi^-|$, R2' and R2*, not R2 and QSM all demonstrated significant differences between these two groups, with p = 0.0010, 0.0022, 0.0091, 0.011, 0.10 and 0.31, respectively.

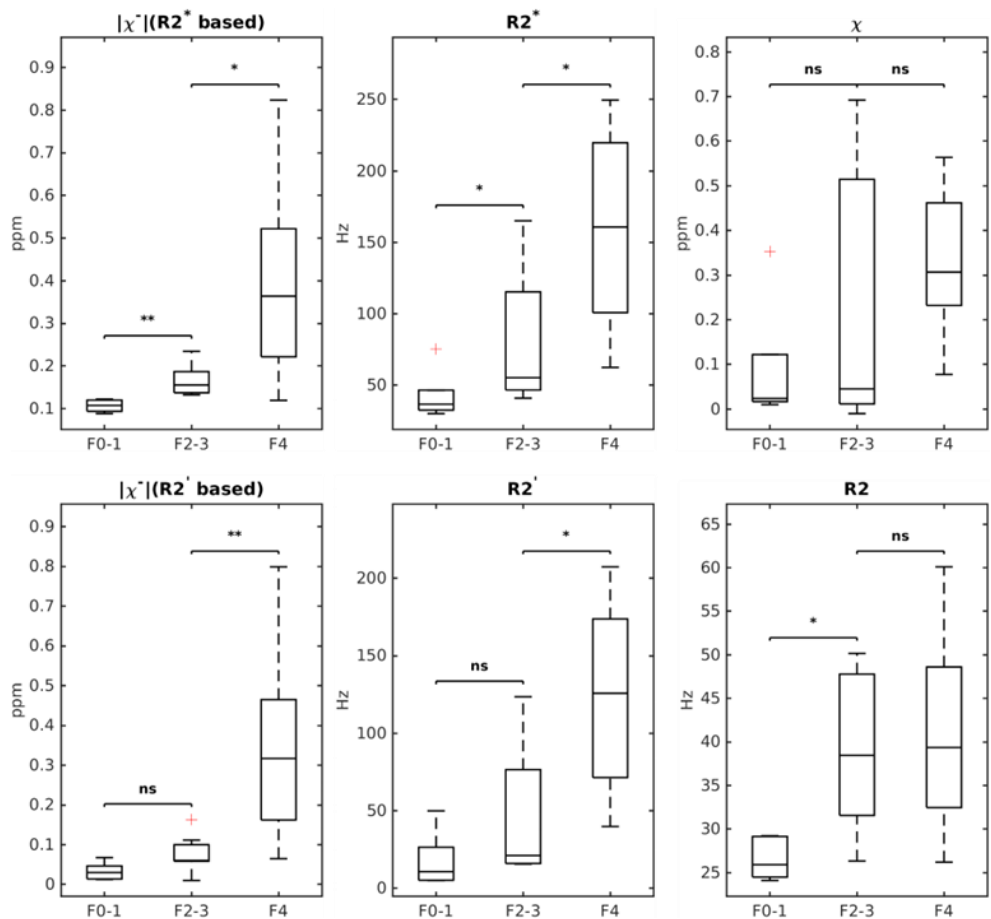

Figure 3: Boxplots of R2*-based $|\chi^-|$, R2'-based $|\chi^-|$, R2*, $\chi$, R2 and R2' values in samples with stages F0-1, F2-3 and F4 respectively. The P-value range of a Mann–Whitney U test comparing the groups is displayed on top of the groups (ns: non-significant, *: 0.01<p<0.05, **: 0.001<p<0.01).

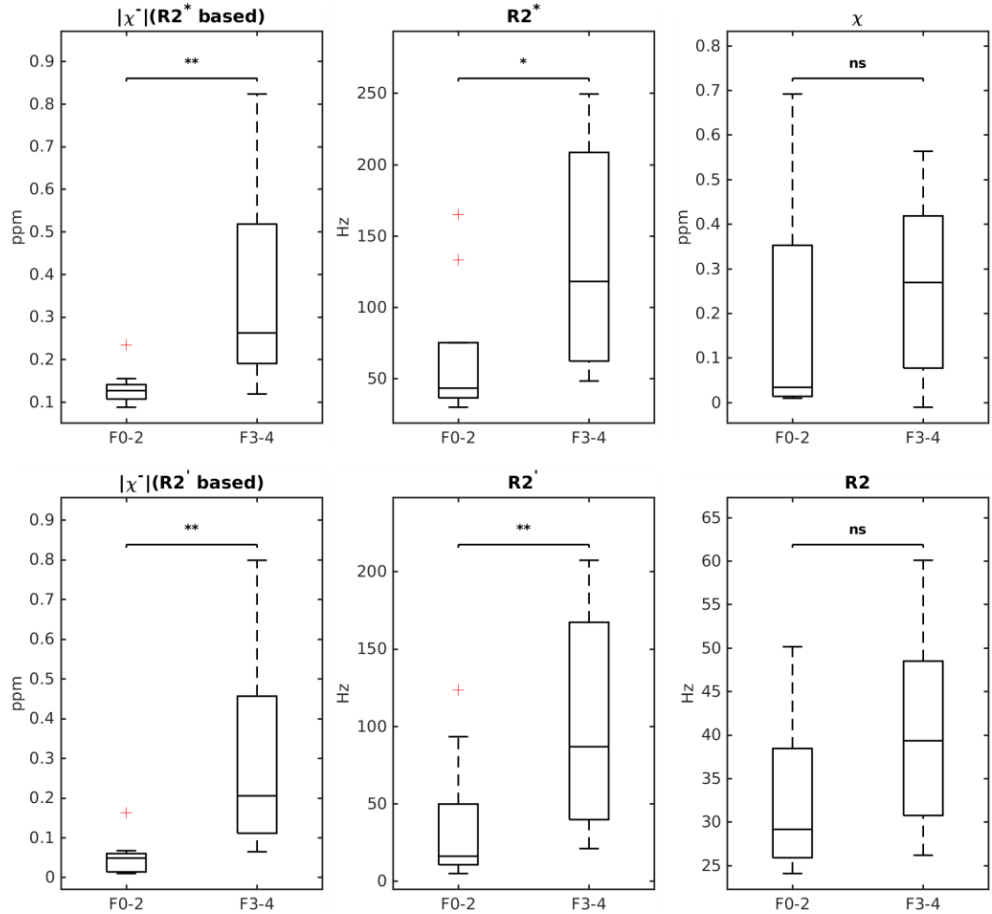

Figure 4: Boxplots of R2* and R2 based $|\chi^-|$, R2*, $\chi$, R2 and R2' values in samples with stages F0-2 and F3-4. The P-value range of a Mann–Whitney U test comparing the groups is displayed on top of the groups (ns: non-significant, *: 0.01<p<0.05, **: 0.001<p<0.01).

## 3.3 | ROC analysis

Figure 5 displays Receiver Operating Characteristic (ROC) curves for various fibrosis differentiations between subgroups of liver samples with stage F0-1 (n=5) and stage F2-4 (n=15). R2* based $|\chi^-|$ enhanced the Area Under the Curve (AUC) to 0.987 (P = 0.0017), indicating a more accurate prediction than when R2* and QSM are used individually. Besides, R2, R2*, R2' and R2'-based $|\chi^-|$ achieved AUCs = 0.947, 0.920, 0.880 and 0.867 and p = 0.0039, 0.0068, 0.015 and 0.018, respectively, for identifying significant stage fibrosis/cirrhosis livers. Magnetic susceptibility ($\chi$) did not reach statistical significance, with AUC of 0.733, and p = 0.14. Additional metrics from the ROC analysis, such as sensitivity and specificity, are detailed in Table 2.

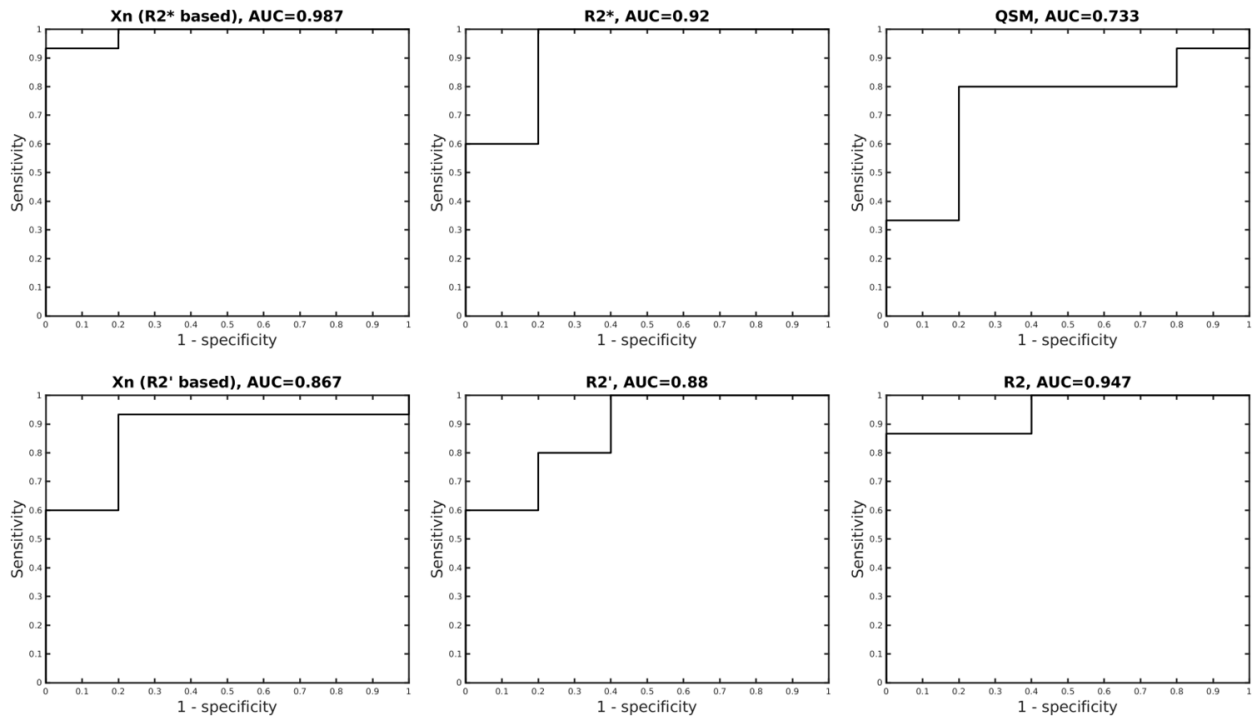

Figure 5: ROC curves for differentiation between samples in non-fibrotic or lower stage fibrotic F0-1 vs. samples with advanced fibrosis or cirrhosis F2-4 for R2*-based $|\chi^-|$, R2'-based $|\chi^-|$, R2*, $\chi$, R2 and R2'.

**TABLE 2** Results of ROC analysis of MRI parameters for differentiation between liver samples with lower stage fibrosis (F0-1, n =5) and advanced liver fibrosis/cirrhosis (F2-4, n = 15).

| Parameters | AUC | P | Cut-off | Sensitivity (%) | Specificity (%) |
|---|---|---|---|---|---|
| $|\chi^-|$ (R2* based) (ppm) | 0.987 | 0.0017 | 0.14 | 93 | 100 |
| R2 (s$^{-1}$) | 0.947 | 0.0039 | 38.55 | 87 | 100 |
| R2* (s$^{-1}$) | 0.920 | 0.0068 | 165.10 | 100 | 80 |
| R2' (s$^{-1}$) | 0.880 | 0.015 | 85.60 | 80 | 80 |
| $|\chi^-|$ (R2' based) (ppm) | 0.867 | 0.018 | 0.19 | 93 | 80 |
| $\chi$ (ppm) | 0.733 | 0.14 | 0.29 | 80 | 80 |

## 4 | DISCUSSION

Our results indicate that R2* based negative susceptibility can serve as an accurate biomarker for assessing liver fibrosis, particularly to differentiate livers with fibrosis F0-1 from those with ≥F2 for antifibrotic pharmacological treatment[42]. Liver's susceptibility sources include diamagnetic fibrosis (prevalently collagen) and paramagnetic iron. In diseases such as nonalcoholic fatty liver disease (NAFLD) and steatohepatitis (NASH), fat is commonly present alongside liver fibrosis. To mitigate the impact of fat on QSM, we used water-fat separation to differentiate water and fat signals and adjust for fat content. To reduce the effect of paramagnetic sources like iron, we used susceptibility source separation to identify the diamagnetic susceptibility source for non-invasively assessing liver fibrosis using ex vivo liver explant samples. While QSM, R2 and R2' individually were not effective in identifying all groups of fibrosis stages, this study found that susceptibility source separation utilizing QSM and R2* from a single mGRE sequence is able to distinguish stage 0-1 from stage 2-3 ($p = 0.0025$) and stage 2-3 from cirrhotic/stage 4 ($p = 0.021$) in liver tissue, providing a ROC AUC of 0.987 for differentiating liver samples of stage F0-1 and advanced stage F2-4 fibrosis/cirrhosis.

This study also compared the performance of R2' based negative susceptibility with R2* based negative susceptibility. Although R2'-based source separation demonstrated slightly better ability in differentiating higher stage fibrosis (F4 vs F2-3 or F3-4 vs F0-2), it failed to distinguish between F0-1 and F2-3, possibly due to large relative scan errors in R2 and R2* maps in lower stage or non-fibrotic livers with smaller R2 and R2* values. Another constraint that limits the application of R2' based source separation in practice is the long acquisition time of mSE sequence for evaluating the R2 map. When acquiring mSE in patients, the requirement of breath-holding may only allow the acquisition of a single slice at a time, and the challenging estimation of R2 map for a 3D volume in clinical practice would require motion compensation strategy such as navigator[43]. On the contrary, 3D mGRE can be acquired in a single breath-hold for a whole liver scan[44].

This study faced several limitations. First, a notably small sample size could have affected the results. To validate these findings and assess their applicability in clinical environments, further studies are necessary, involving larger participant groups. Second, when using the assumption R2'$\approx \alpha$R2*, we estimated $\alpha \approx 0.76$ by fitting the model to our liver data. However, the (R2*, R2') plot in figure 1 exhibits slight deviations from the model when R2* is small or large. The deviation at small R2* may be caused by registration errors between 3D mGRE and 2D m mSE scans; the deviation at large R2* may be caused by additional other factors such as the over-simplified assumptions in the physical models. In the future, a more precise R2' and R2* with various amount of susceptibility sources may be explored. Third, both R2' and R2* based source separation involve substantial simplification and require further investigation. Previous phantom studies demonstrates the strong nonlinear effects of fibrosis and fat interference on the R2* estimate[45] and of signal dependence on voxel size[46]. In the dephasing constant determination, Eqs.3&4 were only approximate, and Eq.6 was only solved approximately. Formalin fixation may cause R2* increase[47,48], and may have affected the evaluation of susceptibility sources and other MRI parameters in our study. Finally, the reproducibility of the results on different scanners and in-vivo using the relaxometry values determined in this paper should be further studied.

In conclusion, our study shows that the diamagnetic susceptibility source obtained from mGRE data alone is a promising tool for noninvasive diagnose of the liver fibrosis stage, which may be used as an alternative of liver biopsy.

**ACKNOWLEDGMENTS**

This research was supported in part by National Institutes of Health (NIH) grants R01CA181566, R01NS090464, R01DK116126 and R01NS095562.

**ORCID**

Gary M. Brittenham https://orcid.org/0000-0002-5667-7446

Yi Wang https://orcid.org/0000-0003-0706-8553